\documentclass[doublecol]{epl2} 

\newcommand{\beq}{\begin{eqnarray}}
\newcommand{\benu}{\begin{enumerate}}
\newcommand{\enu}{\end{enumerate}}
\newcommand{\eeq}{\end{eqnarray}}
\usepackage[latin1]{inputenc}
\usepackage{graphicx,amssymb,amsmath}
\usepackage{indentfirst}
\usepackage{color}
\newcommand{\be}{\begin{equation}}
\newcommand{\ee}{\end{equation}}
\newcommand{\bea}{\begin{eqnarray}}
\newcommand{\eea}{\end{eqnarray}}
\newcommand{\hf}{\frac{1}{2}}

\newcommand{\phib}{\stackrel{\neg}{\phi}}
\newcommand{\lambdab}{\stackrel{\neg}{\lambda}}
\newcommand{\xib}{\stackrel{\neg}{\xi}}

\title{A framework to a mass dimension one fermionic sigma model}
\shorttitle{A framework to a mass dimension one fermionic sigma model} 

\author{R. J. Bueno Rogerio\inst{1}  J. M. Hoff da Silva\inst{1} S. H. Pereira\inst{1} \and Rold\~ao da Rocha\inst{2}}

\institute{      
\inst{1} Departamento de Fisica e Quimica, Universidade
Estadual Paulista, Av. Dr. Ariberto Pereira da Cunha, 333, Guaratinguet\'a, SP,
Brazil.\\
  \inst{2} Centro de Matem\'atica, Computa\c c\~ao e Cogni\c c\~ao, Universidade Federal do ABC 09210-580, Santo Andr\'e,  Brazil.
}

\abstract{ In this paper a mass dimension one fermionic sigma model, realized by the eigenspinors of the charge conjugation operator with dual helicity (Elko spinors), is developed. Such spinors are chosen as a specific realization of mass dimension one spinors, wherein the non-commutative fermionic feature is ruled by torsion. 
 Moreover, we analyse Elko spinors as a source of matter in a background in expansion. {}{Moreover, we analyse Elko spinors as a source
of matter in a background in expansion and we have found that such kind of mass dimension one fermions can serve not only as dark matter but they also induce an effective cosmological constant.}}

\begin{document}

\maketitle
\section{Introduction}

The concept of nonlinear sigma models has been extensively studied in the scientific literature in a broader context. The usefulness of such framework is widely evinced from the symmetries and interconnections among different areas that it allows. Usually, apart from outstanding applications in string theory (see \cite{ST} and references therein), non-linear sigma models have  served to evidence the interplay between  target spaces and Utiyama-Yang-Mills theories \cite{GM}. 

A more subtle issue concerning sigma models is the underlying framework to relativistic fermions. The first order Dirac equation is certainly an additional element of difficulty. The description of the physical reality based upon matter (fermionic fields) and its interactions mediated by abelian and non-abelian gauge theories, however, has motivated the study of possible sigma models in the broader context of spinor theory \cite{CR}. In Ref. \cite{CR}, by exploring the so-called bispinor geometry associated to the bispinor algebra, it was shown that for a particular class of spinors --- whose density and pseudo-scalar density are non-null --- the geometry of the physical observables space is given by a three-dimensional hyperbolic Robertson-Walker space.   

Nearly ten years ago, the systematic study of Majorana spinors has been leading to the appreciation of mass dimension one spinor fields \cite{AHL,Ahluwalia:2009rh}, called Elko\footnote{Eigenspinors of the charge conjugation operator with dual helicity \cite{AHL}.}. This field satisfies the Klein-Gordon, but not the Dirac equation. As a crucial property, these spinors are neutral, under local gauge interactions, by means of the requirement that they are eigenspinors of the charge conjugation operator. In fact, Elko interactions, with matter and gauge fields of the standard model, are suppressed by at least one order of magnitude regarding the unification scale, providing an \emph{ab initio} origin of ``darkness'' of dark matter. In other words, interactions of Elko spinor fields are constrained  to the Higgs field and gravitons, supplying a prominent  direction towards physics beyond the Standard Model.

The theoretical formulation of completeness for these spinors is encoded in the (sub)groups (of the  Lorentz group), which retain the underlying  relativistic structure \cite{JHEPVSR}. Starting from the usual concept of Dirac spinors as elements that carry the representation $(1/2,0)\oplus(0,1/2)$ of the Lorentz group, one can relate the different sectors of the representation space by means of the parity operator $P$. In the context of the full Lorentz group, $P$ is a discrete symmetry and its implementation in a given spinorial formulation culminates in the standard Dirac dynamical equation \cite{LOH}. Nevertheless, the two parts of the representation space can be also related through the so-called ``Pauli matrices magic'' \cite{RAM} without reference to any discrete symmetry. In this context, the resulting dynamics is not provided by the Dirac equation, but rather by the Klein-Gordon equation {}{only}. Working out the particularities of such spinor field in the second quantisation program, a violation of the full Lorentz symmetry appears in the spin sums. Interestingly enough, it is possible to show that the spin sums, and therefore the whole formulation, is invariant under SIM(2) group transformations \cite{JHEPVSR}: precisely the subgroups of the Lorentz group obtained by \;removing the discrete symmetries \cite{COGL}. Additionally, Elko can be experimentally produced by Higgs interactions  \cite{marcao,Agarwal}. 
3M\par
{}{In the standard model of particle physics, all regarded  spinors are Dirac, Weyl or Majorana. Such spinors obey a first-order derivative field equation. This characteristic implies a quantum propagator that, for large momentum, is proportional to $p^{-1}$.
This asymptotic behaviour of the associated propagator results, among other things, in the fact that mass dimension must be $3/2$. Now, the unique kinematic operator that is satisfied by Elko is the Klein-Gordon equation, a second-order derivative field equation. For this case, for a large momentum, the quantum propagator is proportional to $1/p^{2}$, contradicting the previous case. The Klein-Gordon operator is the proper kinetic operator for Elko fields. We are, therefore, led to conclude that the mass dimension of the Elko field is \emph{one}, rather than $3/2$, as it would be usually expected for a fermionic field.}

All the above mentioned physical aspects of these fields (mass dimension one, neutrality, spin 1/2) enable Elko spinors to be dark matter candidates, constructed from very first principles. Hence the systematic investigation of these dark matter candidates, slightly and safely departing from the usual quantum description has led us to the analysis of the mass dimension one fermionic sigma model presented here.   
\par
In this paper we construct and investigate a non-linear sigma model associated to mass dimension one spinor fields. Recent formal results point to the fact that there {}{are many kinds of such spinors} \cite{NOI}. However, to fix ideas,  we shall report to Elko spinors, which are prototypes of mass dimension one fermionic fields. By sigma model, we mean the mapping of the Minkowski space into a complete target space, performed by spinor  fields, and its relationship with Utiyama-Yang-Mills theory. In this vein, bearing in mind the Grassmannian character of the spinor variable, we further endow the target space with torsion. 
We organise this paper as follows: in the next section we depict the general set up of nonlinear sigma model and its relation to non-abelian gauge theories, also endowed to torsion in order to encompass the non-commutative fermionic aspect. Moving forward, in Section III we construct a representative sigma model for mass dimension one fermions. In the final section we conclude. Whenever it is possible, we provide some starting points to the application of the general formulation in cosmology. 

\section{The general setup with torsion}

This section shall present a straightforward generalisation, regarding the interplay between nonlinear sigma models and non-abelian theories \cite{MUK}, encoding torsion terms. As remarked in the Introduction, we envision further applications to a specific, although essentially fermionic, case.  

We start, by depicting the general aspects, not particularising to mass dimension one fermionic fields immediately. Let $\{\xi_i\}$ be the canonical basis of a natural inertial frame in the target space $\Sigma$, and $\{d\xi^i\}$ its dual basis. It is possible to split the target space geometry by defining an effective metric $g\in \sec(T_p\Sigma)^*\times\sec(T_p\Sigma)^*$, where, as usual, $p$ is an arbitrary point belonging to $\Sigma$, and $\sec T_p\Sigma$ is a section of the tangent bundle of $\Sigma$ at $p$, such that given $\phi =\varphi^i\xi_i$, it yields  
\begin{eqnarray}
g(\varphi,\varphi)=[g_{mn}d\xi^m\otimes d\xi^n+\gamma_{mn}d\xi^m\wedge d\xi^n](\varphi^i\xi_i,\varphi^j\xi_j),\label{1}
\end{eqnarray} being $\otimes$ and $\wedge$ the tensor and the exterior product, respectively. The geometrical {}{splitting} is, indeed, fulfilled by Eq. (\ref{1}). A direct computation of (\ref{1}), taking into account the anti-symmetry relation between the two products, leads to 
\begin{eqnarray}
g(\varphi,\varphi)=g_{ij}\varphi^i\varphi^j+\frac{1}{2}\tilde{\gamma}_{ij}\varphi^i\varphi^j,\label{2}
\end{eqnarray} where $\tilde{\gamma}_{ij}=\gamma_{ij}-\gamma_{ji}$. By writing a given product as its commuting and anti-commuting counterparts, i. e. $\varphi^i\varphi^j=\frac{1}{2}[\varphi^i,\varphi^j]+\frac{1}{2}\{\varphi^i,\varphi^j\}$, it yields   
\begin{eqnarray}
g(\varphi,\varphi)=\frac{1}{2}g_{ij}\{\varphi^i,\varphi^j\}+\frac{1}{4}\tilde{\gamma}_{ij}[\varphi^i,\varphi^j].\label{3}
\end{eqnarray} It is worth to mention that a thorough classification of spinors based upon bilinear covariants in a space-time wherein the metric has both symmetric and anti-symmetric parts have been accomplished in \cite{Ablamowicz:2014rpa}.

A complete sigma model, in the sense of Eq. (3), can be thus studied,  by means of the free Lagrangian 
\begin{equation}
\mathcal{L}\!=\!\frac{1}{2}g(\partial_\mu\varphi,\partial^\mu\varphi)\!=\!\frac{1}{4}g_{ij}\{\partial_\mu\varphi^i,\partial^\mu\varphi^j\}+\frac{1}{8}\tilde{\gamma}_{ij}[\partial_\mu\varphi^i,\partial^\mu\varphi^j],\label{4}
\end{equation}
where Greek {}{indexes stand} for space-time coordinates. Obviously, in the usual commutative case, it yields $[\partial_\mu\varphi^i,\partial^\mu\varphi^j]=0$, hence we are simply left with $\mathcal{L}=\frac{1}{2}g_{ij}\partial_\mu\varphi^i\partial^\mu\varphi^j$. In this last case, the connection with Utiyama-Yang-Mills theories is determined by the requirement $\frac{\delta\mathcal{L}}{\delta\varphi_i}=0$ (under space-time volume integration). In this vein, the Christoffel symbols $\Gamma^i_{jk}(\varphi)$ are automatically generated, in terms of which the connection $A_{\mu k}^{\;i}(\varphi)=\Gamma^i_{jk}(\varphi)\partial_\mu\varphi^j$ is identified. The equation of motion, then, reads
\begin{equation}
D_{\mu \;j}^{\;i}(\partial^\mu\varphi^j)=0,\label{6}
\end{equation} 
where the covariant derivative is given by $D_{\mu \;j}^{\;i}=\partial_\mu\delta_j^i+A^{\;i}_{\mu j}$. Besides, the contraction of the Riemann curvature tensor $R^i_{jkl}$ with $\partial_\mu\varphi^k\partial_\nu\varphi^l$ leads to the non-abelian field strength
\begin{eqnarray}
R^i_{jkl}(\partial_\mu\varphi^k)(\partial_\nu\varphi^l)=\partial_\mu A^{\;i}_{\nu j}-\partial_\nu A^{\;i}_{\mu j}+([A_\mu,A_\nu])^i_{\; j}.\label{7}
\end{eqnarray}

Returning to the complete case, including the non-commutative sector, the functional variation of the Lagrangian leads to
\begin{eqnarray}
\partial_\mu\partial^\mu\varphi^m+\tilde{\Gamma}^m_{ij}(\varphi)\partial_\mu\varphi^i\partial^\mu\varphi^j=0,\label{tor1}
\end{eqnarray} 
with
\begin{eqnarray}
\tilde{\Gamma}^m_{ij}(\varphi) = \Gamma^m_{ij}(\varphi) + \Lambda^{m}_{ij}(\varphi)\,,\label{gammatilde}
\end{eqnarray}
where $\Lambda^{m}_{ij}(\varphi)$ is defined as
\begin{eqnarray}
\Lambda^{m}_{ij}(\varphi) \equiv \frac{1}{2}g^{mk}\Big(\partial_{i}g_{jk} - \partial_{j}g_{ik} - \frac{1}{2}\partial_{k}\tilde{\gamma}_{ij}\Big),
\end{eqnarray}
and we have the explicit contribution of the torsion terms. Moreover, the general target space curvature tensor is given by 
\begin{eqnarray}
\tilde{R}^i_{jkl} = R^{i}_{jkl} + \partial_{[k}\Lambda^{i}_{jl]} + (\Gamma^{i}_{m[k}+\Lambda^{i}_{m[k})\Lambda^{m}_{jl]} + \Lambda^{i}_{m[k}\Gamma^{m}_{jl]}.\label{tor2}
\end{eqnarray} 
Notice that, if $\tilde{\gamma}_{ij}=0$, and requiring $\left[\partial_{\mu}\varphi^{i},\partial^{\mu}\varphi^{j}\right]=0$, (culminating with $\Lambda^{m}_{ij}(\varphi) = 0$), Eqs. (\ref{tor1}) and (\ref{tor2}) reduce to the usual case, as expected. Besides, as a matter of fact, it is not trivial (and, perhaps, not even insightful) to find the Yang-Mills counterpart of the geometric quantities as in (\ref{7}),  for the case at hand. The important aspect to be stressed here is that,  when commutativity is lifted in the target space, torsion terms are generated.  

In order to envisage the implementation of an application, we depict some cosmological implications of the sigma model. We shall briefly present the main equations that must be considered, given a specific form for the target space fields.  In a curved background the action containing the corresponding contribution of the sigma model Lagrangian (\ref{4}) reads \be
S=\int\sqrt{-\bar{g}}d^4x\Bigg[ -\frac{R}{2\kappa} + \hf h_{ij}\partial_\mu\varphi^i \partial_\nu\varphi^j \bar{g}^{\mu\nu} - W(\varphi)\Bigg]\,,\label{action}
\ee
where $R$ is the Ricci scalar, $\kappa=8\pi G$ and $h_{ij}=g_{ij}+\hf \tilde{\gamma}_{ij}$ can be obtained from the symmetric and anti-symmetric property of (\ref{3}). The space-time metric is represented here by $\bar{g}^{\mu\nu}$ and $W(\varphi)$ stands for a self-interacting potential.

The variation of action (\ref{action}), with respect to the metric $\bar{g}^{\mu\nu}$, leads to the Einstein equations
\be
R_{\mu\nu}-\hf \bar{g}_{\mu\nu}R = 2\kappa T_{\mu\nu}\,,\label{Einstein}
\ee
where $R_{\mu\nu}$ is the Ricci tensor, and $T_{\mu\nu}$ is the canonical energy-momentum tensor corresponding to the matter content:
\be
\!\!\!T_{\mu\nu}=h_{ij}\partial_\mu\varphi^i \partial_\nu\varphi^j -  \bar{g}_{\mu\nu}[ \hf h_{ij}\partial_\alpha\varphi^i \partial_\beta\varphi^j\bar{g}^{\alpha\beta} - W(\varphi)]\,.\label{EM}
\ee
Variation with respect to $\varphi^k$ leads to the equation of motion for the fields
\be
\frac{1}{\sqrt{-\bar{g}}}\partial^\mu(\sqrt{-\bar{g}} \partial_\mu \varphi^m)+\tilde{\Gamma}^m_{ij}(\partial_\mu\varphi^i\partial^\mu\varphi^j)+(\partial_k W)g^{km}=0\,,\label{eq_motion}
\ee
with $\tilde{\Gamma}^m_{ij}$ given by (\ref{gammatilde}).

It is straightforward to check that (\ref{eq_motion}) reduces to (\ref{tor1}), when the space-time metric $\bar{g}^{\mu\nu}$ is the Minkowski flat metric and the potential is null. By specifying the fields $\varphi^i$, and the potential $W(\varphi)$, we can obtain the Friedmann equations from (\ref{Einstein}). An emergent universe,  supported by a non-linear sigma model without torsion,  was studied in \cite{chervon}. In the context approached here, by taking advantage of the fact that $h_{ij}$ encodes torsion terms, it is quite plausible that the dynamics of the evolution will be changed. 

\section{Building up the sigma model}

Part of the structure of Elko spinors, $\lambda$, is built upon the requirement $C\lambda=\pm \lambda$ being $C$ the charge conjugation operator. There exists the self conjugated spinors {}{$\lambda^{S}_{\alpha}$ $(C\lambda^S_{\alpha}=+\lambda^S_{\alpha})$ and the anti-self conjugated spinors $\lambda^A_{\alpha}$ $(C\lambda^A_{\alpha}=-\lambda^A_{\alpha})$. Moreover, a quite judicious analysis shows that the right dual to $\lambda$ (from the relativistic point of view) reads $\lambdab_\alpha=\pm i[\lambda_\beta]^\dagger \gamma^0$ \cite{DUAL}, where the labels $\alpha$ and $\beta$ denote different types of spinors and, clearly, the dual relation stands for both self and anti-self conjugated spinors. As a last necessary remark, we remember that there are four different Elko spinors: two of them corresponding to different states for the self conjugated case, and similarly to the anti-self conjugated case \cite{AHL}.}

{}{It is possible to provide a particular basis adapted to eigenspinors of the charge conjugation operator, related to 
the Majorana representation. Alternatively, we can use the  chiral representation, paradigmatically explored in all the literature of Elko. Whatever the basis is chosen, it is worth to mention that our approach is basis independent. }
It is useful for our purposes to construct all the possible spinors as follows: let $c^i$ and $d_j$ be $c$-numbers and write $\lambda=c^i\lambda_i$ and $\lambdab=d_j\lambdab^j$. Now let us decompose $\lambda$ (and $\lambdab$) in terms of the usual canonical basis (and its corresponding dual basis) as 
\begin{eqnarray}
\lambda_i=\left(\begin{array}{cc}
                                \lambda_i^1\\
                                \lambda_i^2 \\
					\lambda_i^3\\
					\lambda_i^4\\
                              \end{array}
                            \right)=\lambda_i^1\left(\begin{array}{cc}
                                1\\
                                0 \\
					0\\
					0\\
                              \end{array}
                            \right)+\cdots+\lambda_i^4\left(\begin{array}{cc}
                                0\\
                                0\\
					0\\
					1\\
                              \end{array}
                            \right),\label{8}
\end{eqnarray} 
\begin{eqnarray} 
\lambdab^j&=&\left(\lambdab_1^j,\; \lambdab_2^j,\; \lambdab_3^j,\; \lambdab_4^j \right)=\lambdab_1^j(1,\; 0,\; 0, \;0)+\cdots \nonumber\\&&+\lambdab_4^j(0,\; 0,\; 0, \;1).\label{9}
\end{eqnarray} Hence, by denoting the element basis by $\{\xi_a\}$ (and the corresponding dual by $\{\xib^b\}$), we have $\lambda=c^i\lambda_i^a\xi_a$ and $\lambdab=d_j\lambdab^j_b\xib^b$, with $\xib^b\!\!\!(\xi_a)=\delta^b_a$. In order to properly implement the sigma model to the case at hand, it is necessary to modify the tensor and exterior products, encompassing the fermionic character of the fields. 

We start by defining the product $\tilde{\otimes}$ in the following way: let $\mathbb{S}$ be the complex vector space generated by all the finite linear combinations of usual Cartesian products $(v_i,\stackrel{\neg}{v}^i)$, where $v_i$ and $\stackrel{\neg}{v}^i$ are spanned by the respective canonical basis. Besides, take $\mathbb{I}$ as the subspace of $\mathbb{S}$ generated by 
\begin{eqnarray}
(v_i+u_i,\stackrel{\neg}{w}^i)-(v_i,\stackrel{\neg}{v}^i)-(u_i,\stackrel{\neg}{v}^i),\label{10}\\ (v_i,\stackrel{\neg}{w}^i+\stackrel{\neg}{u}^i)-(v_i,\stackrel{\neg}{u}^i),\label{11}\\(k v_i,l \stackrel{\neg}{w}^i)-kl(v_i,\stackrel{\neg}{w}^i),\label{12}
\end{eqnarray} being $k,l$ c-numbers and $v_i$, $u_i$, and so on, spanned by means of the canonical basis (similarly for the dual case). The product $\tilde{\otimes}$ is defined by conjugating elements in the space $\mathbb{S}/\mathbb{I}$, and therefore Eqs. (\ref{10})-(\ref{12}) ensure bilinearity. Hereupon, we shall pinpoint some important remarks in order to clarify the relevant properties of $\tilde{\otimes}$. Obviously, a basis of $\mathbb{S}/\mathbb{I}$ is given by $\{\xi_i\tilde{\otimes}\xib^i\}$. We define the action of $\tilde{\otimes}$ as 
\begin{eqnarray}
(v\tilde{\otimes}\stackrel{\neg}{w})(\stackrel{\neg}{u},x)=(v^i\xi_i\tilde{\otimes}\stackrel{\neg}{w}_j\xib^j)(\stackrel{\neg}{u}_k\stackrel{\neg}{\xi}^k,x^l\xi_l)=v^i\stackrel{\neg}{w}_j\stackrel{\neg}{u}_ix^j,\label{13}
\end{eqnarray} where $\stackrel{\neg}{w}_j$ and $\stackrel{\neg}{u}_j$ are just coefficients and, more importantly, we have defined the action on the basis as $(\xi_i\tilde{\otimes}\xib^j)(\xib^k,\xi_l)=\xib^k\!\!(\xi_i)\xib^j\!\!(\xi_l)=\delta^k_i\delta^j_l$. It is important to stress that the product $\tilde{\otimes}$ is unique, up to isomorphisms\footnote{From a complementary point of view, the defined product $\tilde{\otimes}$ is just an isomorphism of the usual tensor product $\otimes$.}. Finally, let $\{\xi_i\}\cup\{\xi_i\tilde{\otimes}\xib^j\}$ over the field $\mathbb{C}$ be the basis of $\tilde{\mathbb{T}}$ and $\tilde{\mathbb{I}}$ the bilateral ideal generated by $\xi_i\tilde{\otimes}\xib^i$. The product $\tilde{\wedge}$ acting on $\tilde{\mathbb{T}}/\tilde{\mathbb{I}}$ is related to $\tilde{\otimes}$ is, as expected, given by $\xi_m\tilde{\wedge}\xib^n=\frac{1}{2}(\xi_m\tilde{\otimes}\xib^n\!\!-\;\xi_n\tilde{\otimes}\xib^m)$. 

We are now in position to implement the splitting of Eq. (\ref{1}), although this time endowed with the tilde products. Hence it yields  
\begin{eqnarray}
\mathcal{L}=\frac{1}{2}g^m_n\xi_m\tilde{\otimes}\xib^n\!\!(\partial^\mu\lambdab,\partial_\mu\lambda)\!+\!\frac{1}{2}\gamma^m_n\xi_m\tilde{\wedge}\xib^n(\partial^\mu\lambdab,\partial_\mu\lambda).\label{14}
\end{eqnarray} According to our previous construction, the fermionic decomposition along with the bilinearity of tilde products  can be used  in a fairly direct fashion. It is necessary, however, to call attention to the fact that the Fermi-Dirac statistics is a key feature of Elko formulation \cite{AHL}, which one cannot preclude. Therefore in the adopted decomposition $\lambda=c^i\lambda_i^a\xi_a$ and $\lambdab=d_j\lambdab^j_b\xib^b$ the terms $\lambda_i^a$ and $\lambdab^j_b$ are understood as Grassmannian variables\footnote{Roughly speaking this is the attempt to reproduce quantum features of the field by absorbing the creation/annihilation operators into the expansion coefficients.}. Taking advantage of these remarks, the Lagrangian (\ref{14}) yields 
\begin{eqnarray}
\mathcal{L}=\frac{1}{2}g^i_j d_a c^b\partial^\mu\lambdab^a_i\partial_\mu\lambda^j_b+\frac{1}{4}\tilde{\gamma}^i_j d_a c^b\partial^\mu\lambdab^a_i\partial_\mu\lambda^j_b\,.\label{15}
\end{eqnarray} {}{It is important to stress that the formal structure of the Elko spinors is constructed taking advantage of the spinor rest frame [4]. Therefore, it is quite conceivable to add a mass term in the above Lagrangian. Therefore, it reads
\begin{eqnarray}
\mathcal{L}&=&\frac{1}{2}g^i_j d_a c^b\partial^\mu\lambdab^a_i\partial_\mu\lambda^j_b+\frac{1}{4}\tilde{\gamma}^i_j d_a c^b\partial^\mu\lambdab^a_i\partial_\mu\lambda^j_b\nonumber\\&& + m^2d_{j}c^{i}\lambdab^{j}_{a}\lambda^{a}_{i}. \label{massa}
\end{eqnarray}} It serves as the starting point for the formulation of a mass dimension one fermionic sigma model. Generally speaking, the target space is undertaken as a coset space of the isometry by the isotropy group. In relevant cases  torsion can be added to such coset spaces \cite{PLB}. The Lagrangian (\ref{massa}) can be adjusted to encompass these situations, $\{\xi_a\}$ and $\{\xib^b\}$ being the bases in the fiber bundle formulation, accordingly. From the Lagrangian density (\ref{massa}), it is interesting to note that the torsion terms do affect the spin current density. In fact, the appearance of $\tilde{\gamma}^i_j$ terms in the expression below  makes this point explicit:
\begin{equation}
\!\!S^\mu_{\alpha\beta}=-\frac{1}{2}d_ac^b(g^i_j+\frac{1}{2}\tilde{\gamma}^i_j)\Bigg[\partial^\mu\stackrel{\neg}{\lambda}^{a}_{i}\frac{\delta\lambda^j_b}{\delta \omega^{\alpha\beta}}+\frac{\delta\stackrel{\neg}{\lambda}^{a}_{i}}{\delta\omega_{\alpha\beta}}\partial^\mu\lambda^j_b\Bigg]. \label{spd}
\end{equation}

We shall finalize by considering the Elko Lagrangian (\ref{massa}) as the source of matter, in a curved expanding background. Some care is necessary, in order to correctly write the corresponding action in such a case. First, we define the fields $\phib_i=d_a \lambdab_i^a$, $\phi^j=c^b \lambda^j_b$, and introduce the covariant derivatives, by $\nabla_\mu \phib_i \;= \partial_\mu \phib_i + \phib_i \Gamma_\mu$,  and $\nabla_\mu \phi^j = \partial_\mu \phi^j -\Gamma_\mu \phi^j$, where $\Gamma_\mu$ are the spin connections coupling Elko spinors to the background metric. The action in a curved background reads 
\be
S=\int\sqrt{-\bar{g}}d^4x\Bigg[ -\frac{R}{2\kappa} + \hf h^i_j\nabla_\mu\phib_i \nabla_\nu\phi^j \bar{g}^{\mu\nu} - W(\phib,\phi)\Bigg]\,,\label{action2}
\ee
where $h^i_{j}=g^i_{j}+\hf \tilde{\gamma}^i_{j}$. The equations of motion for the fields follow  directly, by taking the variation with respect to $\phib_i$ and $\phi^j$. Eq. (\ref{action2}) may serve as the starting point to apply the formulation presented here in the cosmological context. In particular, the quartic interaction appearing due to the (spin connection) torsion contribution (as in the usual fermionic case) is generated. 

{}{In order to show an explicit contribution coming from the symmetric and anti-symmetric part of the sigma model target space into the cosmological equations, let us take for simplicity a set of constant Elko spinor fields, such that $\partial_\mu\lambdab \,=0=\partial^\mu\lambda$, and the potential as the quadratic one in the form $W={1\over 2}m^2 \lambdab \lambda$. We can choose to write $h^i_j$ in the following simple form:
\beq
h^i_j=\left(\begin{array}{cccc}
g & \gamma & \gamma & \gamma  \\
-\gamma & g & \gamma & \gamma\\
-\gamma & -\gamma & g & \gamma \\
-\gamma & -\gamma & -\gamma & g
\end{array} \right)\,,
\eeq
with $g$ and $\gamma$ representing the symmetric and anti-symmetric components of $h^i_j$. It may sound as an oversimplification, however (as we shall see) this particularization leads to an relevant physical consequence. In a flat, homogeneous and isotropic FRW metric, $\bar{g}_{\mu\nu}=diag [1,\,-a(t)^2,\,-a(t)^2,\,-a(t)^2]$ the spin connections $\Gamma_\mu$ can be determined as $\Gamma_0=0$ and $\Gamma_k=-{\dot{a}\over 2}\gamma^0\gamma^k$, where $\gamma^\mu$ are the standard Dirac matrices and a dot stands for time derivative. Notice that, even for constant spinor fields,  the spin connection term couples to the metric, through the term $\phib_i\Gamma_\mu\Gamma^\mu \phi^j$, with $\Gamma_\mu\Gamma^\mu=-{3\over 4}{\dot{a}^2\over a^2}\mathbb{I}$ and, therefore, contributes with a time dependent term. The action (\ref{action2}) can be written as:
\be
S=\int\sqrt{-\bar{g}}d^4x\Bigg[ -\frac{R}{2\kappa} + \Lambda_S(t) + \Lambda_A(t) - \Lambda_m \Bigg]\,.\label{action3}
\ee
The $\Lambda_m$ term comes from the potential part, namely $\Lambda_m={1\over 2}m^2\lambdab \lambda$, and represents a cosmological constant term, with a dependence on the mass of the Elko spinor field. The terms $\Lambda_S(t)$ and $\Lambda_A(t)$ stands for the symmetric and anti-symmetric contributions coming from the sigma model respectively, and can be written as $\Lambda_S(t)={3\over 8} g C(d_a,c^b,\lambdab^a_i,\lambda_b^i)H(t)^2$ and $\Lambda_A(t)={3\over 16}\gamma C(d_a,c^b,\lambdab^a_i,\lambda_b^i)H(t)^2$, where $C(d_a,c^b,\lambdab^a_i,\lambda_b^i)$ is a constant and $H(t)=\dot{a}/a$ is the Hubble parameter. They act as an effective time varying cosmological constant, $\Lambda_{eff}(t)=\Lambda_S(t) + \Lambda_A(t)-\Lambda_m$. At early times of cosmological evolution, when $\Lambda_S(t) + \Lambda_A(t)>\Lambda_m$, the positive contribution from $\Lambda_{eff}$ acts as an attractive gravitational field in the geometric side of the Einstein equation, leading to a decelerating universe. As the universe expands $H(t)$ decreases. When the condition $\Lambda_S(t) + \Lambda_A(t) < \Lambda_m$ is reached the universe turns to be dominated by a negative $\Lambda_{eff}$, which implies a repulsive gravitational force, driven by a constant cosmological term $\Lambda_m$, in perfect agreement to the $\Lambda CDM$ model. It is straightforward to see that the constant parameters could be adjusted, in order to reproduce the transition from a decelerated to an accelerated expansion of the universe. A much more rich scenery concerns the study of Elko spinor fields dynamic coupled to the gravitational field \cite{BOE6}.}


\section{Concluding remarks}

We analyzed and studied a mass dimension one fermionic sigma model, realized by Elko spinors. A non-commutative fermionic feature was introduced by the prominent role of torsion. The effective connection (\ref{gammatilde}) that rules the Euler-Lagrange equations (\ref{tor1}) is defined with respect to the anti-symmetric part of the metric in the target space. Thereat, Elko spinors play the role of a source of matter in an expanding background. 

By the very nature of mass dimension one fermions, the study of the action (\ref{action2}) concerning the sigma model for Elko spinors can provide further insights on the dark matter problem. However, it is also interesting to pursue questions concerning the sigma model itself. For instance, the study of the non-minimal coupling between the Riemann tensor and four Elkos, as in the standard realization of supersymmetric sigma models. Moreover, the usual $N=1$ (D=4) supersymmetric sigma model case also imposes the necessity of a Kaehlerian target-space. Hence the investigation of what type of geometric condition may arise from the extension of the present work to the supersymmetric case is also in order. We shall delve these questions in the future.


\section{Acknowledgments}
The authors express their gratitude to professor Jos\'e Abdalla Helayel-Neto for the privilege of his comments and appreciation of the manuscript. JMHS thanks to CNPq grants No. 308623/2012-6 and No. 445385/2014-6 for partial support. RdR is grateful for the CNPq grants No. 303293/2015-2 and No. 473326/2013-2, and FAPESP grant No. 2015/10270-0, and to INFN grant ``Classification of Spinors'', which has provided partial support. SHP thanks to CNPq grants No. 304297/2015-1.


\begin{thebibliography}{99}

\bibitem{ST} \Name{Tseytlin A. A.} \REVIEW{Int. J. Mod. Phys. A}{4}{1989}{1257}.

\bibitem{GM} \Name{Gell-Mann M. \and Levy M.} \REVIEW{Nuovo Cimento}{16}{1960}{705}. 

\bibitem{CR} \Name{Crawford J. P.} \REVIEW{J. Math. Phys.}{31}{1990}{1991}.

\bibitem{AHL}      \Name{Ahluwalia D. V. \and Grumiller D.} \REVIEW{JCAP}{0507}{2005}{012}.

\bibitem{Ahluwalia:2009rh}
  \Name{Ahluwalia D. V., Lee C. Y. \and Schritt D.} \REVIEW{Phys.\ Rev.\ D}{83}{2011}{065017}.
                  
\bibitem{JHEPVSR}  \Name{Ahluwalia D. V. \and  Horvath S. P.} \REVIEW{JHEP}{1011}{2010}{078}. 


\bibitem{LOH} \Name{Speran\c ca L. D.} \REVIEW{Int. J. Mod. Phys. D}{23}{2014}{1444003}.

{}{\bibitem{RAM} \Name{Ramond P.} {\it Field Theory: A Modern Primer}, Second Edition, Westview Press (2001).}

\bibitem{COGL} \Name{Cohen A. G.  \and  Glashow S. L.} \REVIEW{Phys. Rev. Lett.}{97}{2006}{021601}.


\bibitem{marcao} \Name{Dias M., de Campos F. \and  Hoff da Silva J. M.} 
\REVIEW{Phys. Lett. B}{706}{2012}{352}.

\bibitem{Agarwal} 
  \Name{Agarwal B., Jain P., Mitra S., Nayak A.~C. \and Verma R.~K.}
\REVIEW{Phys.\ Rev.\ D}{92}{2015}{075027}.


{}{\bibitem{NOI} \Name{Coronado Villalobos C. H.,  Hoff da Silva J. M. \and da Rocha R.} \REVIEW{Eur. Phys. J. C}{75}{2015}{266}.}

{}{\bibitem{MUK} \Name{Mukhi S. {\it Classical and Quantum Theory of Supersymmetric $\sigma$-Models} in {\it Superstrings, Unified Theories and Cosmology}, Summer Workshop in High Energy Physics and Cosmology, Trieste, Italy (1986); Edited by: G. Furlan, R. Jengo, J. C. Pati, D. W. Sciama, E. Sezgin, Q. Shafi. }}


\bibitem{Ablamowicz:2014rpa} 
  \Name{Ab\l amowicz R., Gon\c calves I. \and ~da Rocha R.}
  \REVIEW{J.\ Math.\ Phys.}{55}{2014}{103501}.


\bibitem{chervon} \Name{Beeshan A.,  Chervon S. V. \and  Maharaj S. D.} \REVIEW{Class. Quant. Grav.}{26}{2009}{075017}.

{}{\bibitem{DUAL} \Name{Ahluwalia D. V., Lee C. Y., \and Schritt D.} \REVIEW{Phys.\ Lett.\ B}{687}{2010}{248}.}



{}{\bibitem{PLB} \Name{Batakis N. A., Farakos K.,  Kapetanakis D.,  Koutsoumbas G. \and  Zoupanos G.} \REVIEW{Phys. Lett. B}{220}{1989}{513}.  }

{}{\bibitem{BOE6} \Name{B\"ohmer C. G.,  Burnett J.,  Mota D. F. \and  Shaw D. J.}
\REVIEW{JHEP}{07}{2010}{053}.}
\end{thebibliography}
\end{document}